\def\mytitle{Unique Games on the Hypercube}
\def\myauthors{Naman Agarwal\footnote{nagarwl3@illinois.edu, University of
Illinois Urbana-Champaign}, Guy Kindler\footnote{gkindler@cs.huji.ac.il,
Hebrew university, Jerusalem, Israel. Supported by grants from the
Israel Science Foundation and the Binational Science Foundation.}, Alexandra
Kolla\footnote{akolla@illinois.edu, University of Illinois Urbana-Champaign},
Luca Trevisan\footnote{trevisan@stanford.edu, Stanford University}}
\def\mybib{Hypercubes}

\documentclass[11pt]{article}
\usepackage[fullpage,nousetoc,nohylinks]{paper}
\theoremstyle{plain}
\newtheorem*{theorem*}{Theorem}
\newtheorem*{lemma:main_lemma}{Lemma \ref{lemma:main_lemma}}
\newtheorem*{lemma:gap_preservation}{Lemma \ref{lemma:gap_preservation}}
\usepackage{hyperref}
\BEGINDOC

\makeheader

\begin{abstract} In this paper, we investigate the validity of the Unique Games Conjecture when the constraint graph is the boolean hypercube. We construct an almost optimal integrality gap instance on the Hypercube for the Goemans-Williamson semidefinite program (SDP) for Max-2-LIN$(\Z_2)$. We conjecture that adding triangle inequalities to the SDP provides a polynomial time algorithm to solve Unique Games on the hypercube.
\end{abstract}

\section{Introduction}
\newcommand{\remove}[1]{}

\newcommand{\maxlin}{Max-2-LIN$(\Z_2)$ }
\def\MC{Max-Cut}
\def\SC{Sparsest-Cut}
\newcommand{\gwplus}{GW+}

The \textit{Unique Games Conjecture} (UGC) was formulated by Khot in
2002~\cite{Kho02}, and has since been the focus of great attention.
Deciding it either way would have significant implications. Proving
the conjecture would give tight bounds to the approximability of
several fundamental optimization problems, including {Vertex
  Cover}~\cite{KhotRegev}, {\MC}~\cite{KKMO} and non-uniform
{\SC}~\cite{CKKRS,KV}). Refuting the conjecture would yield 
an approximation algorithm for finding small non-expanding sets \cite{RS10},
and the techniques used in a refutation would be likely to find applications
to other graph partitioning problems like \MC\ and Sparsest-Cut.

\begin{conjecture}(UGC) For any constants $\epsilon, \delta > 0$,
  there is a $k=k(\epsilon,\delta)$ such that it is NP-hard to
  distinguish between instances of Unique Games with alphabet size $k$
  where at least $1-\epsilon$ fraction of constraints are satisfiable
  and those where at most $\delta$ fraction of constraints are
  satisfiable. 
\end{conjecture}

A full definition of a Unique Games instance appears in
Section~\ref{sec:prelim}, but for now one may think of an instance
with alphabet size $k$ as a system of constraints where variables take
values in $\Z_k$, and each constraint is a linear equation $\bmod k$ involving two variables.
The {\em constraint graph} of such an instance is a graph that has a vertex for every variable
and an edge for every pair of variables that appear together in one of the constraints.

\medskip 

In recent years the UGC was found to be intimately connected to the
power of semi-definite programming (SDP). One can trace the first instance
of this connection to the seminal paper by Goemans and Williamson~\cite{GW94} on the
Max-Cut problem (an instance of Max-Cut can be thought of as a linear
equation system over $\Z_2$, and thus it is a Unique Games Instance
for alphabet size $2$). They gave an SDP based algorithm for MaxCut
which, on inputs where the maximal cut is of size $(1-\epsilon)$,
produces a cut that satisfies at least $(1 - (2/\pi)\sqrt{\epsilon})$
fraction of the constraints. A matching integrality gap was found by
\cite{Car} and \cite{FS}, and in \cite{KKMO} it was proven that if
the UGC is correct, than the Goemans-Williamson algorithm is the polynomial-time
approximation algorithm with the best
possible approximation ratio. 

\begin{theorem}\label{thm_KKMO} 
  Assume the
  Unique Games Conjecture. Then for all sufficiently small $\epsilon >
  0$, it is NP-hard to distinguish instances of \maxlin that are at
  least $(1 - \epsilon)$-satisfiable from instances that are at most
  $(1 - (2/\pi)\sqrt{\epsilon})$-satisfiable.  
\end{theorem}

The Goemans-Williamson SDP algorithm was later extended to general
Unique Games by \cite{CMM1,CMM06}, and the approximation ratio
achieved by \cite{CMM06} is shown to be tight in
\cite{KKMO}. Raghavendra~\cite{Rag08} proved that for every constraint
satisfaction problem there is a polynomial-time, semi-definite
programming based algorithm which, if UGC is true, achieves the best
possible approximation for the problem.

\medskip There is very limited evidence in favor or against the Unique
Games conjecture. In the direction of trying to refute the conjecture, 
the most promising algorithmic approaches rely on semidefinite programming relaxations. Various
semidefinite programs and rounding schemes have been studied, among which
are~\cite{Kho02,Trevisan,GT,CMM1,CMM2}. Integrality gap results have
been sought to understand the limitation of current approaches and as
partial evidence of the hardness of unique games. Integrality gap
instances have been constructed for the basic relaxation in \cite{KV,
  RS09,
  BGHMRS, KPS}, and these instances show that the basic relaxation, and certain
extensions of it, cannot be used to refute the UGC. A polynomial-time
solvable extension of the basic relaxation, however, solves
near-optimally all the known integrality gap instances \cite{BBHKSZ},
and it is a possible candidate for an algorithm refuting the UGC.
In~\cite{ABS}, a sub-exponential ($2^{n^{\Omega(1)}}$-time) algorithm for general instances
is given, based on spectral techniques.

Unique games are known to have good polynomial-time or quasi-polynomial-time  spectral
approximation algorithms for large classes of instances. These include
expanders~\cite{AKKTSV,MM}, local expanders~\cite{AIMS,RS10}, and more
generally, graphs with few large eigenvalues~\cite{K}. Spectral algorithms also solve
nearly-optimally, in time at most $2^{n^{o(1)}}$, all the known inegrality
gap instances mentioned above (here $n$ denotes the number of vertices of the constraint graph), which, 
as mentioned, are also well approximated in polynomial time by the SDP studied in  \cite{BBHKSZ}. Thus, such instances are known to not be ``hard'' for Unique Games.  We note that instances generated in
various semi-random models~\cite{KMM} are solvable in polynomial time.

Algorithmic work on restricted classes of instances can lead to
progress toward the resolution of the UGC, whether it ultimately
proves to be true or false. If the UGC is false, then the algorithmic
work on restricted classes of instances represents steady progress
toward its refutation. Indeed, the algorithmic breakthrough of
\cite{ABS} came by combining a way of dealing with instances in which
the constraint graph has few large eigenvalues, following \cite{K},
with a new way to deal with the complementary case of instances in
which the constraint graph has several large eigenvalues. If the UGC
is true, then the best evidence we can hope for in the short term is
the discovery of integrality gaps for polynomial-time solvable
relaxations, including the relaxations studied in \cite{BBHKSZ}.
Moreover, the algorithmic work on restricted classes of instances, by
identifying which instances are easy, can be used to guide the search
for hard families of instances.

In order to make algorithmic progress on UGC, the question to ask next
is what type of graphs are the ones on which known techniques have failed to
yield polynomial (or at least $2^{n^{o(1)}}$-time) provably good
approximation. As a consequence of its expansion profile, the hypercube is a worse-case instance for the known stat-of-the-art spectral algorithms, which require time $2^{n^{\Omega(1)}}$ to solve Unique Games on it. Moreover, as far as the authors are aware, there has been no results until this work on how various SDP relaxations perform on the hypercube.

The purpose of this work is to understand the Unique Games problem when
the input constraint graph is the hypercube. Since the hypercube
is a good representative of the ``last frontier'' instances on which
the Unique Games problem is still not known to be easy, finding an efficient algorithm that solves UG on the hypercube might give
some motivation to suspect that UGC is false. On the other hand,
constructing integrality gap instances in which the constraint graph
is a hypercube present certain unique difficulties, and requires
constructions of a different nature from the ones that have been
developed so far \cite{KV, RS09, BGHMRS, KPS}. 

An integrality gap instance for
a relaxation of an optimization problem is an instance for which the optimum of the relaxation is $\geq 1-\epsilon$, while the
optimum of the  problem is $\leq 1-\epsilon'$, for some $\epsilon<< \epsilon'$; in 
all the previous integrality gap instances of unique games (and also integrality gap instances of max cut and other constraint satisfaction problem), the feasible solution witnessing that the optimum of the relaxation is $\geq 1-\epsilon$ is constructed in such a way that every constraint contributes $\geq 1-\epsilon$ to the cost function. In a unique game on the hypercube, however, if the instance is unsatisfiable then there is an unsatisfiable subset of just four constraints, since for an unsatisfiable instance there need to be four-cycles that are inconsistent. Thus, there cannot be a feasible solution for a relaxation in which every edge contributes more than $3/4$ to the cost function. Being forced to reason about non-symmetric solutions might give new ideas that could be applied in more general settings.

\subsection{Our results}
In this paper, we consider  the \maxlin problem (recall that, by theorem \ref{thm_KKMO}, an improved
approximation algorithm for \maxlin\ for general instances would refute the UGC)
and we study its approximability when restricted to instances whose constraint graph is an hypercube.

We construct a family of integrality gap instances of Unique Games on the hypercube constraint graph for the Goemans-Williamson semi-definite program (SDP). In the following, we refer to an edge  of the constraint graph of a \maxlin instance as an \textit{equality} or an \textit{inequality} edge, depending on whether the constraint on the edge is satisfied when its endpoints have the same or opposite value, respectively.

\paragraph{The GW algorithm.} As mentioned above, the first
non-trivial algorithm for Unique-Games was the SDP based algorithm
introduced by Goemans and Williamson \cite{GW95}. This algorithm was
later generalized by Charikar-Makarychev-Makarychev \cite{CMM1} to
give the best known (worst-case) polynomial time approximation for
Unique Games instances. The Goemans-Williamson (GW) SDP applies to
Unique Games instances with $2$ labels, referred to as \maxlin, and
for an instance which is $1-\epsilon$ satisfiable, the GW algorithm
finds an assignment satisfying $1-O(\sqrt \epsilon)$ fraction of the
constraints. Assuming the UGC, theorem \ref{thm_KKMO} implies that
this performance is best possible by any algorithm. As also mentioned
above, tightly matching integrality gaps were shown in \cite{FS} and
\cite{Car}, and in \cite{FS} instances were constructed where any
reasonable rounding procedure for the GW SDP gives an approximation
factor matching the above parameters.

In the context of solving UG on the hypercube, it is natural to ask
how well the GW SDP performs. As far as the authors are aware, until
this work it was not ruled out that the GW algorithm performs (almost)
perfectly, which would imply that instances with a hypercube
constraint graph are easy.

\paragraph{Integrality gap on the hypercube.}
In this paper, we prove that the GW SDP has an integrality gap on the
hypercube with a behaviour similar to the gap on general graphs,
albeit with a different power of $\epsilon$.

\begin{theorem*}(\textbf{Main}) For every sufficiently small constant
  $\epsilon$, and for every $d\geq d(\epsilon)$, there exists a
  \maxlin instance on the boolean cube $Q_{d}$ of dimension $d$ such that the UG combinatorial optimal value for that
  instance is $1- \Omega(\epsilon)$, and the GW SDP optimal value is
  $1-O(\epsilon^{3/2})$. \end{theorem*}

We believe our integrality gap can be extended for the case of more
than $2$ labels, but have no proof of that at this point. We note that
this result is especially interesting since all previously known
integrality gap instances for the GW SDP as well as most integrality
gap instances for other various SDPs, are known to not be ``hard''  instances for
Unique Games and can be approximately solved using spectral
techniques in time at most $2^{n^{o(1)}}$, where $n$ is the number of node of the graph. The hypercube graph is unique in the sense that known state-of-the-art spectral algorithms cannot solve it in time faster than $2^{n^{\Omega(1)}}$, (where $n$ is the number of nodes of the hypercube) and, as shown by this work, it also provides an integrality gap instance for the GW SDP.


\paragraph{Adding triangle inequalities.} Adding so called ``triangle
inequality constraints'' is a standard manipulation of semidefinite
programs. We show that adding these constraints to the GW SDP breaks
the integrality gap of our instance. We conjecture that in fact the GW
SDP with triangle inequalities solves the Unique Games problem on the
hypercube.

\paragraph{Our Techniques.}
We construct our gap instance by starting with an instance for which all edges
are equality edges and converting a small number edges to inequalities ensuring
that the all-one's assignment is still roughly the combinatorial optimum
assignment, while at the same time the SDP optimum decreases.
More concretely we show the following lemma :

\begin{lemma:main_lemma}[Main Lemma]
\label{lemma:main_lemma}
For every sufficiently large $d$, there exists a \maxlin Instance on the
hypercube $Q_d$ of dimension $d$, such that the combinatorial optimum is $1 -
\Omega(d^{-1/2})$ and the SDP optimum is $1 - \Oh(d^{-3/4})$.
\end{lemma:main_lemma}

In the following, we refer to edges of the hypercube connecting vertices that differ in the $i$-th coordinate as edges {\em going in the $i$-th direction}.

To prove the above lemma we define a gap instance $\Delta(k,d)$ on $Q_d$, where
we start from the all-equalities instance on $Q_d$ and introduce inequalities
along $k$ directions, for $k \sim \sqrt{d}$. Our goal in choosing which edges to designate
as inequality edges is to keep the solutions in which all variables are assigned the same value (say, the value one) to be
close to optimal, which implies that the combinatorial optimum is
roughly one minus the fraction of inequality edges, while
at the same time allowing an SDP solution of value noticeably higher than one minus the fraction of inequalities, thus creating a gap.

We show that if we restrict ourselves to introducing inequality edges in just one direction, then we can show that up to about half the edges going in that direction can be changed to inequality while preserving the property that the all-one assignment to the variables is optimal, and while allowing an SDP solution of higher cost.

Next, we further extend the construction by placing these inequality regions along $\Oh(\sqrt{d})$ number of directions. The parameter
$\Oh(\sqrt{d})$ is chosen such that the all ones assignment is
still nearly the optimum. In particular in Lemmas \ref{lemma:comboptimum} and
\ref{lemma:sdpoptimum} we prove that if we place these inequality regions in $k$ directions, the
combinatorial optimum grows linearly with $k$ (in particular $\Oh(\frac{k}{d})$)
whereas the SDP optimum grows at most proportionally to $\sqrt{k}$ (in particular
$\Oh(\frac{\sqrt{k}}{d})$). Setting $k \sim \Oh(\sqrt{d})$ we get a
non trivial (super-constant) gap. We formally state and prove the above idea in
lemma \ref{lemma:main_lemma}. 

So far, we have managed to show that a non trivial gap instance exists for
sub-constant $\epsilon \sim d^{-1/2}$. The next task is to blow this instance up
to create gap instances for constant $\epsilon$. We do this by showing the following gap preservation lemma:

\begin{lemma:gap_preservation}[Gap Preservation]
\label{lemma:gap_preservation}
Suppose that $I$ is a \maxlin instance define over the $d$
dimensional hypercube, and let $\alpha$ be the combinatorial value of
$I$ and $\beta$ be the optimal SDP value for it. Then for every $i$
there exists an instance $I'$ defined over the $d\cdot i$ dimensional
hypercube whose combinatorial value is at least $\alpha$ and whose GW
SDP optimal value is at most $\beta$.
\end{lemma:gap_preservation}

We prove the above lemma by defining a \textit{tensor} product operation on
\maxlin instances on the cube which allows us to create larger instances of
the cube preserving the gap of the original instance.

\remove{\paragraph{Open Questions.}
next step: integrality gap for hierarchies. this dons't work with our
instance: we prove that it has disjoint cycle certificates, so even
sdp with triangle inequalities gives correct number on our instance. 

open: whether sdp with triangle inequalities works for all 2-label
instances on the cube}

\subsection{Organization}
The rest of the paper is organized as follows. Section
\ref{sec:prelim} contains some definitions and notation that we will
be using throughout the paper. Sections \ref{sec:mainthm} and
\ref{sec:gappreservation} contain the description of the integrality
gap instance on the hypercube and the proof of our Main theorem. In
section \ref{sec:cycles} we discuss adding triangle inequalities to
the GW algorithm, show that our instance no longer gives an
integrality gap for  this strengthened SDP, and conjecture that
actually this SDP solves all instances on the hypercube. 


\section{Preliminaries}\label{sec:prelim}
\subsection{Notations}
We use the following notations throughout the paper.  $Q_{d}$ refers to a
hypercube graph $(V_d,E_d)$ of dimension $d$, with vertex set $V_d$ and 
edge set $E_d$. We generally reserve $d$ for the dimension of the hypercube in context. 
For every vertex $v \in V_d$ of the
hypercube we naturally associate vector $\mathbf{v} \in \{0,1\}^d$ (we
denote vectors in boldface letters). Let $\mathbf{v}_i$ be the
$i^{th}$ coordinate of the vector $\mathbf{v}$. We denote by $H(\mathbf{v})$ the hamming
weight of the vector $\mathbf{v}$, i.e.
the number of $1'$s in $\mathbf{v}$.

Let $Q_{d-k}(\mathbf{x})$ where $\mathbf{x} \in \{0,1\}^k$ be a $d-k$
dimensional sub-cube of of $Q_d$ obtained by fixing the first $k$ coordinates to
be $\mathbf{x}$.

Given any two vectors we define their tensor product as follows 

\begin{definition}[Tensor Product]
Given two vectors $\mathbf{x} \in \R^{n_1},\mathbf{y} \in \R^{n_2}$ define the
vector $\mathbf{x} \otimes \mathbf{y} \in \{0,1\}^{n_1*n_2}$ as follows

$$\mathbf{x} \otimes \mathbf{y}_{(i*n_1 + j)} = \mathbf{x}_i \mathbf{y}_j$$

It will well known that the norm of tensor products is multiplicative,

\begin{itemize}
  \item $\|\mathbf{x} \otimes \mathbf{y}\|^2 = \|x\|^2\|y\|^2$,
\end{itemize} 

where $\|\mathbf{x}\|$ denotes the $l_2$-norm of $\mathbf{x}$.
\end{definition}

\subsection{Unique Games and \maxlin Definitions}

Following is a formal genera definition of the Unique Games problem.

\begin{definition}[Unique Games]
  A \textit{Unique Games} instance for alphabet size $k$ is specified
  by an undirected constraint graph $G = (V,E)$, a set of variables
  $\{x_u\}_{u \in V}$, one for each vertex $u$, and a set of
  permutations (constraints) $\pi_{uv}: [k] \rightarrow [k]$, one for
  each $(u,v)$ s.t.\ $\{u,v\}\in E$, with $\pi_{uv} =
  (\pi_{vu})^{-1}$.  An assignment of values in $[k]$ to the variables
  is said to satisfy the constraint on the edge $\{u,v\}$ if
  $\pi_{uv}(x_u) = x_v$.  The optimization problem is to assign a
  value in $[k]$ to each variable $x_u$ so as to maximize the number
  of satisfied constraints. We define the \textit{value} of a solution
  to be the fraction of constraints that are \textit{not} satisfied by
  this solution.
\end{definition}

An optimal solution for a Unique Games instance which satisfies the maximum
number of constraints will also be referred to as the \textit{combinatorial
solution} and its value will be referred to as the \textit{combinatorial value}.
We note that, while it is slightly more common to define the value of a solution
to be the fraction of satisfied constraints by the solution, in this paper we
find it more convenient to define the value of a solution as the fraction of
constraints that is not satisfied by it.

\begin{definition}[Max-2-Lin]
A \maxlin instance is a Unique Games instance with alphabet size 2. Note that to specify a \maxlin instance it
is sufficient to specify a graph $G=(V,E)$ and a function $f : E
\rightarrow \{0,1\}$. We call the edges with the value 0 ``equality'' edges and
the edges with value 1 ``inequality'' edges in accordance to the constraints
implying whether the two labels on the edge should be equal or not.

Since the main focus of this paper are instances where the constraint graphs are
Hypercubes, we will be viewing a \maxlin instance as a function $I:E_d
\rightarrow \{0,1\}$. 
\end{definition}

\subsection{Goemans Williamson SDP and Gap Instances}

We next describe the Goemans-Williamson semi-definite program \cite{GW95} for
solving \maxlin. Note that the original paper by Goemans Williamson defines
the SDP for the Max-Cut problem which is essentially a \maxlin problem with
all edges being inequality edges.

\begin{definition}[GW SDP]
Given a graph $G=(V,E)$ and a \maxlin instance $I:E \rightarrow \{0,1\}$ on
it, let the set of equality edges be $E^{+}$ and the set of inequality edges be
$E^{-}$. The Goemans-Williamson SDP for the instance is defined as 
\begin{alignat*}{2}
    \text{minimize }   & \frac{1}{4|E|}\left( \sum\limits_{(u,v) \in E^{+}}
    \|x_u - x_v\|^2 +  \sum\limits_{(u,v) \in E^{-}} \|x_u +
    x_v\|^2 \right)
    \\
    \\
    &\text{subject to } \;\; \|x_u\|^2 = 1 \;\; (\forall \; u \in V)
  \end{alignat*}
\end{definition}

Note that the analysis carried out by Goemans-Williamson for Max-Cut essentially
holds for \maxlin  too. In particular given a \maxlin instance with combinatorial
value $\epsilon$ the above SDP has optimum value $\Omega(\epsilon^2)$. As mentioned in the introduction, this is
tight under the Unique Games Conjecture.


\section{Main Theorem}\label{sec:mainthm}
In this section, we describe an instance of \maxlin on the Hypercube which is an
integrality gap for the GW SDP. 

\begin{definition}[$(\alpha,\beta$)-gap instance]
  An infinite family $\mathcal F$ of \maxlin instances on the
  hypercube (of varying dimensions) is called {\em an
    $(\alpha,\beta)$-gap Instance} for \maxlin if the combinatorial
  optimum on any $I\in\mathcal F$ is $\Omega(\alpha)$ and the GW SDP
  has optimum value $\Oh(\beta)$.
\end{definition}

Note that the existence of an $(\alpha,\beta)$-gap instance for
\maxlin proves that the GW SDP has an integrality gap of at least
$\Omega(\frac{\alpha}{\beta})$. The following theorem therefore
establishes an integrality gap for the GW SDP on the hypercube.

\begin{theorem}(Main)\label{thm:main}
For every sufficiently small constant $\epsilon$ there is an
$(\epsilon,\epsilon^{3/2})$ Instance for \maxlin.
\end{theorem}

Theorem~\ref{thm:main} is obtained from the following  two lemmas,
which are proven below.

\begin{lemma}(Main Lemma)
\label{lemma:main_lemma}
For every sufficiently large $d$, there exists a \maxlin instance
defined over the $d$ dimensional hypercube, whose combinatorial value
is $\Omega(d^{-1/2})$ but for which the GW SDP optimal value is
$O(d^{-3/4})$.
\end{lemma}

\begin{lemma}[Gap Preservation]
\label{lemma:gap_preservation}
Suppose that $I$ is a \maxlin instance define over the $d$
dimensional hypercube, and let $\alpha$ be the combinatorial value of
$I$ and $\beta$ be the optimal GW SDP value for it. Then for every $i$
there exists an instance $I'$ defined over the $d\cdot i$ dimensional
hypercube whose combinatorial value is at least $\alpha$ and whose GW
SDP optimal value is at most $\beta$.
\end{lemma}

\begin{proof}[Proof of Theorem~\ref{thm:main}]
  Let $\epsilon>0$ be small enough, and take $d = 1/\epsilon^2$.  By
  lemma $\ref{lemma:main_lemma}$ we can find an instance $I$ whose
  combinaorial value is at least $\Omega(d^{-1/2})\sim\epsilon$ and
  whose GW SDP optimal value if $\Omega(d^{-3/4})\sim
  \epsilon^{3/2}$. Considering the family of instances that can be
  obtained from $I$ by applying Lemma~\ref{lemma:gap_preservation} we
  obtain a  $(\epsilon,\epsilon^{3/2})$ Instance of \maxlin, proving
  the theorem.
\end{proof}

In the rest of this section we prove
Lemma~\ref{lemma:main_lemma}. Lemma~\ref{lemma:gap_preservation} is
proven in the next section.

\subsection{Proof of Lemma \ref{lemma:main_lemma}}

We need to construct a \maxlin instance over the hypercube $Q_s$ of
dimension $d$. Let $E_d$ be the set of edges of the hypercube, and let
$k$ be a parameter to be fixed later. We define the instance
$\Delta[k,d]:E_d \rightarrow \{0,1\}$ as follows. For any edge
$e=(v_1,v_2)$ let $i(e)$ be the coordinate along which the
corresponding vectors $\mathbf{v_1},\mathbf{v_2}$ differ. Let
$H(\mathbf{v}[k])$ be the hamming weight of the vector $\mathbf{v}$
restricted to only coordinates other than the first $k$ coordinates.

\begin{itemize}
  \item If $i(e) > k$, $\Delta[k,d](e)=0$.
  \item if $i(e) \leq k$ and if $H(\mathbf{v_1}[k]) > \frac{d-k}{2}$,
  $\Delta[k,d](e)=0$.
  \item Otherwise $\Delta[k,d](e)=1$. 
\end{itemize}

We now make some observations about our instance. Note that all edges that are
assigned to 1 (i.e. are inequality edges) are between vertices $(v,v')$ that differ in one of the first $k$ coordinates.
Therefore for any subcube $Q_{d-k}(\mathbf{x})$ defined by fixing the first $k$
coordinates to be
$\mathbf{x}$, we have that the edges inside $Q_{d-k}(\mathbf{x})$ are all set to
0 (i.e.
are equality edges). Consider two vectors $\mathbf{x_1},\mathbf{x_2} \in \{0,1\}^k$
which differ in one coordinate. Every vertex $v$ in the subcube
$Q_{d-k}(\mathbf{x_1})$ is connected by an edge to another vertex $v'$ in
$Q_{d-k}(\mathbf{x_2})$. The vertex $v'$ can be thought of as a copy of $v$ in
the subcube $Q_{d-k}(\mathbf{x_2})$ (restricted to the last $d-k$ coordinates the two vertices are the same). The edge connecting $(v,v')$ is an inequality or equality edge depending on which side of the majority cut $v$ belongs to in its corresponding subcube. Namely, if more than half of the last $d-k$ coordinates of $v$ are 1, then $(v,v')$ is an equality edge, otherwise it is an inequality edge.

We now bound the GW-SDP optimum and the combinatorial optimum of
$\Delta[k,d]$ in the following two lemmas. 

\begin{lemma}
\label{lemma:comboptimum}
For $k \leq O(\sqrt{d})$, $\Delta[k,d]$ has combinatorial optimum
$\Omega(\frac{k}{d})$.
\end{lemma}

\begin{lemma}
\label{lemma:sdpoptimum}
$\Delta[k,d]$ has GW-SDP optimum $O(\frac{\sqrt{k}}{d})$.
\end{lemma}

Lemma \ref{lemma:main_lemma} easily follows from
Lemma~\ref{lemma:comboptimum} and Lemma~\ref{lemma:sdpoptimum} by setting
$k=c\sqrt{d}$. It is thus left to prove the two lemmas above.

\subsection{Proof of Lemma \ref{lemma:sdpoptimum}}

To prove the lemma it is enough to exhibit a valid solution to the GW-SDP which
achieves a value of $O(\frac{\sqrt{k}}{d})$. To this end we exhibit a two
dimensional solution $S:V_d \rightarrow \R^2$. Our
solution will map every vertex $v$ to a unit vector in $\R^2$ and therefore it
is enough to just specify the angles $\alpha_v$ between $v$ and the $x$-axis. 

The solution $S$ is symmetric with respect to the $d-k$ dimensional subcubes
$Q_{d-k}(\mathbf{x})$ and depends only upon the parity of the $k$ dimensional
vector $\mathbf{x}$.
Within a subcube, the vector assigned to a vertex depends only on the hamming
weight of the vertex restricted to the subcube. Let
$L_i(\mathbf{x})$ be the layer in the $d-k$ dimensional subcube
$Q_{d-k}(\mathbf{x})$ of hamming weight $i$ (vertices with $i$ ones in the last
$d-k$ coordinates).
Formally a vertex $v \in L_i(\mathbf{x})$ if $v \in Q_{d-k}(\mathbf{x})$ and
$H(\mathbf{v}[k]) = i$ (as a reminder, $H(\mathbf{v}[k])$ is the hamming weight of the vector
$\mathbf{v}$ restricted to only coordinates after the $k^{th}$ coordinate).

We now define our solution $S$ to the GW-SDP paramterized by $t$. We
will find a suitable value for $t$ when we analyze the value of the solution.

\begin{itemize}
  \item For every $k$-length vector $\mathbf{x^{+}}$ of parity 1, and for all
  $v \in L_i(\mathbf{x^{+}})$
  \begin{equation*}
  \alpha_v = \left\{
	\begin{array}{ll}
		0  & \mbox{if }  i \leq \frac{d-k}{2} - t \\
		\frac{\pi}{4}\left(1 - \frac{\frac{(d-k)}{2} - i}{t}\right) & \mbox{if } i \in
		 (\frac{d-k}{2} - t, \frac{d-k}{2} + t) \\
		\frac{\pi}{2} & \mbox{if } i\geq \frac{d-k}{2} + t
	\end{array}
\right.
  \end{equation*}

   \item For every $k$-length vector $\mathbf{x^{-}}$ of parity $-1$, and for all $v
   \in L_i(\mathbf{x^{-}})$ assign $\alpha_v$ to be $\pi -$ the corresponding
   value for its neighboring vertex $\mathbf{x^{+}}$ of parity 1. i.e.
  \begin{equation*}
  \alpha_v = \left\{
	\begin{array}{ll}
		\pi  & \mbox{if }  i \leq \frac{d-k}{2} - t \\
		\pi - \frac{\pi}{4}\left(1 - \frac{\frac{(d-k)}{2} - i}{t}\right) & \mbox{if }
		i \in (\frac{d-k}{2} - t, \frac{d-k}{2} + t) \\
		\frac{\pi}{2} & \mbox{if } i\geq \frac{d-k}{2} + t
	\end{array}
\right.
  \end{equation*}
   
\end{itemize}

Following is a schematic of the solution described above. $L$ represents
layers of subcubes with parity 1 and the $L'$ represent their counterparts in
subcubes of parity $-1$

\begin{figure}[h!]
\centering
\includegraphics[width=0.9\textwidth]{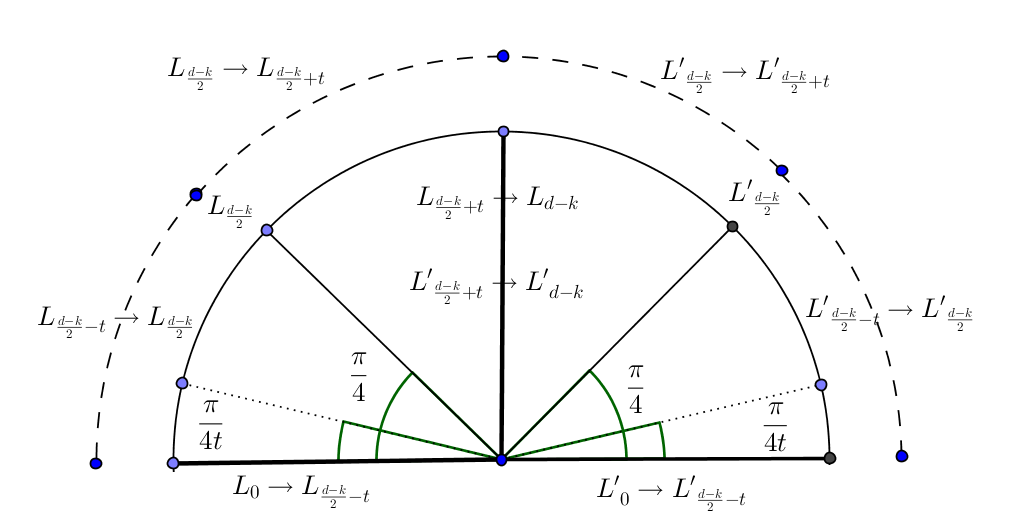}
\caption{Schematic for SDP solution}
\end{figure}

We first compute the contribution of a fixed subcube
$Q_{d-k}(\mathbf{x})$ to the GW-SDP objective. Consider a vertex $v \in L_i(\mathbf{x})$ where $i
\in [0,\frac{(d-k)}{2} - t]$.
This vertex is connected with equalities to its neighbours inside the
subcube and with inequalities to its neighbours outside the subcube. Since all
its neighbours inside the subcube are mapped to the same vector, the contribution to the SDP value
of those edges is zero. Moreover all neighbors of $v$ in different subcubes are mapped to the antipodal point of the vector $v$ is mapped to (since neighboring subcubes have different parity). Therefore the
contribution of every edge connected to this vertex is 0.

Similarly, for a vertex $v \in L_i(\mathbf{x})$ where $i \in [(d-k)/2 +
t,(d-k)]$ the contribution of all its edges is 0. 

Consider a vertex $v \in L_i(\mathbf{x})$ where $i \in ((d-k)/2 - t,(d-k)/2
)$. The total contribution of the neighbors of this vertex comes from the
inequalities going out of the subcube, which is $$k(1 + cos(\pi - 2\alpha_v)) \leq 2k$$
 and from the equalities inside the subcube, which is
$$(d-k)(1-cos(\frac{\pi}{4t}))$$
The total contribution of edges adjacent to $v$ therefore is

\begin{equation*}
2k + (d-k)(1-cos(\frac{\pi}{4t})) \leq \Oh(k + \frac{(d-k)}{t^2})
\end{equation*}

The total fraction of vertices contained in layers $L_i(\mathbf{x})$ for $i
= ((d-k)/2 - t,(d-k)/2 + t)$ is $\Oh(t/\sqrt{d-k})$ (for t=1 it is 
$\theta(\frac{1}{\sqrt{d-k}})$ and that is the layer with the largest fraction
of vertices).
Therefore the total contribution of a fixed subcube $Q_{d-k}(x)$ is bounded by 

$$ |V_{d-k}(x)|\Oh(\frac{t}{\sqrt{d-k}}\left( k + \frac{(d-k)}{t^2} \right) $$

Substituting $t = \sqrt{\frac{d-k}{k}}$ and summing the contribution over all
subcubes $Q_{d-k}(x)$ we get that the fractional value of this SDP feasible solution is
$\Oh(\frac{\sqrt{k}}{d})$.
 
\subsection{Proof of Lemma \ref{lemma:comboptimum}} 

We show here one proof of Lemma~\ref{lemma:comboptimum}. An
alternative proof appears in Theorem~\ref{thm:conjholdsforinstance}:
that proof not just shows a bound on the combinatorial optimum but
also shows that there is a certificate for this bound using
inconsistent cycles.

Consider the $d-k$ dimensional subcubes $Q_{d-k}(\mathbf{x})$
where $\mathbf{x}$ is a $k$ dimensional vector. We first prove that in an
optimum assignment, the assignment on any subcube $Q_{d-k}(\mathbf{x})$ is
determined only by the parity of $\mathbf{x}$. In other words, if
$\mathbf{x},\mathbf{y}$ are $k$ dimensional vectors with the same parity then
the assignments on the subcubes $Q_{d-k}(\mathbf{x}),Q_{d-k}(\mathbf{y})$ will
be the same. We prove this by contradiction. 

Let an optimum assignment be $\Gamma:V_d \rightarrow \{0,1\}$. For a subset of
edges $E \subseteq E_d$ let $Val_{\Gamma}(E)$ be the number of unsatisfied edges
in $E$. Let $Val_{\Gamma}(Q_{d-k}(\mathbf{x}))$ be the number of unsatisfied edges in
the subcube $Q_{d-k}(\mathbf{x})$. 

Let $S$ be the set of pairs of $k$ dimensional vectors
$\mathbf{x_1},\mathbf{x_2}$ which differ in one coordinate. Given any
two such vectors $\mathbf{x_1},\mathbf{x_2}$, let $E(\mathbf{x_1},\mathbf{x_2})$ be the set of
edges $(u,v)$ that go between the subcubes i.e. 
$u \in Q_{d-k}(\mathbf{x_1}),v \in Q_{d-k}(\mathbf{x_2})$. Therefore the total
combinatorial value of the assignment $\Gamma$(i.e. total number of unsatisfied
edges) is

\begin{multline}
\sum\limits_{\mathbf{x}} Val_{\Gamma}(Q_{d-k}(\mathbf{x})) +
\sum\limits_{(\mathbf{x_1},\mathbf{x_2}) \in S}
Val_{\Gamma}(E(\mathbf{x_1},\mathbf{x_2})) 
= \\ \sum\limits_{(\mathbf{x_1},\mathbf{x_2}) \in S} \left( \frac{1}{k}\left(
Val_{\Gamma}(Q_{d-k}(\mathbf{x_1})) + Val_{\Gamma}(Q_{d-k}(\mathbf{x_2}))
\right) + Val_{\Gamma}(E(\mathbf{x_1},\mathbf{x_2})) \right)
\end{multline}

Given the above expression let $\mathbf{x_1'},\mathbf{x_2'}$ be vectors such
that the quantity inside the summation in the RHS above is minimum. Now
consider the assignment in which for every vector $\mathbf{x}$ which has the same parity as
$\mathbf{x_1'}$, the subcube $Q_{d-k}(\mathbf{x})$ has the same assignment as 
the subcube $Q_{d-k}(\mathbf{x_1'})$ in $\Gamma$. We do the same with $\mathbf{x_2'}$. 
It is easy to see that the above described assignment satisfies at least as many
edges as $\Gamma$.

By the above argument for any optimal assignment $\Gamma$ it is enough to
specify two assignment functions $\Gamma_+ : Q_{d-k} \rightarrow \{0,1\}$ and
$\Gamma_- : Q_{d-k} \rightarrow \{0,1\}$, one for subcubes for which the first $k$ coordinates have parity $1$ and one for subcubes for which the first $k$ coordinates have parity $-1$. Let $Val(\Gamma_+)$
and $Val(\Gamma_-)$ be the number of edges not satisfied within the subcubes of positive and negative
parity respectively. Let $Val(\Gamma_+,\Gamma_-)$ denote the number of edges
not satisfied between a fixed subcube of positive parity and a fixed subcube of
negative parity. The total number of edges not satisfied by the assignment
$\Gamma$ therefore is 

$$ 2^{k-1}(Val(\Gamma_+) + Val(\Gamma_-)) + 2^{k-1}k(Val(\Gamma_+,
\Gamma_-))$$

We now prove that without loss of generality the assignment $\Gamma_-$ can be
assumed to be the all 1's assignment $\mathbf{1}$ i.e. $\mathbf{1}(v)=1$ for
all $v \in Q_{d-k}$. 

Consider any optimal assignment $(\Gamma'_+$,$\Gamma'_-)$. Consider the
assignment such that $\Gamma_-=\mathbf{1}$ and $\Gamma_+ = \Gamma'_+ \oplus \Gamma'_-$.
Note that $Val(\Gamma'_+,\Gamma'_-) = Val(\mathbf{1},\Gamma'_+
\oplus \Gamma'_-)$. Also note that $Val(\mathbf{1}) = 0$ and $Val(\Gamma'_+
\oplus \Gamma'_-) \leq Val(\Gamma'_+) + Val(\Gamma'_-)$. Therefore the
assignment $(\mathbf{1},\Gamma'_+ \oplus
\Gamma'_-)$ is at least as good as $(\Gamma'_+,\Gamma'_-)$.

In accordance with the above observations for an optimal assignment it is enough
to specify the assignment $\Gamma_+$ for the positive parity subcubes. 

Consider an optimal solution $(\Gamma'_+,\mathbf{1})$. Let $d'=d-k$. Let $V_{0}
\subseteq V_{d'}$ be the set of vertices $v$ such that $\Gamma_+(v) = 0$ and $H_{1/2}
\in V_{d'}$ be the set of vertices $v$ such that $H(v) \leq d'/2$. Now it is
easy to see that the number of edges unsatisfied by the assignment
$(\Gamma'_+,\mathbf{1})$ is
\begin{eqnarray*}
&&2^{k-1}\biggr(Val(\Gamma_+) + Val(\mathbf{1}) + k*Val(\Gamma_+,
\mathbf{1})\biggr) \\
&& = 2^{k-1} \biggr(E[V_{0},V_{d'}\backslash V_0] + 0 + k(|H_{1/2}-V_0|+|V_0
 - H_{1/2}|)\biggr) \\
 && = 2^{k-1} \biggr(E[V_{0},V_{d'}\backslash V_0] + 0 + k(|H_{1/2}| - |H_{1/2}
 \cap V_0| + |V_0 - H_{1/2}|)\biggr) \\
 && = 2^{k-1} \biggr( \frac{k*2^{d'}}{2} - A(k,d') \biggr)
\end{eqnarray*}

where $A(k,d') = k(|V_0 \cap  H_{1/2}| -
|V_0 - H_{1/2}|) - E[V, V_{d'} - V_0]$. We show in lemma
\ref{lemma:combopthammer} that the above defined quantity $A(k,d') \leq
\frac{k\alpha}{2}*2^{d'}$ for $k \leq \Oh(\sqrt{d})$ where $\alpha$ is a universal constant. 

Therefore the total fraction of edges unsatisfied by the any optimum assignment
is $\Oh(k/d)$ for $k = \Oh(\sqrt{d})$

\begin{lemma}
\label{lemma:combopthammer}
Let $Q_d$ be the hypercube of dimension $d$. Let $V_d$ be the vertex
set of the cube and let $V \subseteq V_d$. Let $H_{1/2}$ be the set of
vertices with hamming weight $\leq d/2$. Let $k \leq
\frac{\pi}{2}\mathcal{I}(Maj_d)$, where $\mathcal{I}(Maj_d)=\Theta(\sqrt
d)$ is the influence of the majority function on $d$ coordinates.  Then

$$A(k,d) \eqdef k(|V \cap  H_{1/2}| -
|V \backslash H_{1/2}|) - E[V, V_d \backslash V] \leq \alpha\frac{k2^{d}}{2}$$ 

where $\alpha$ is a constant $< 1$. 
\end{lemma}
\begin{proof}
Let $Min_d$, $Maj_d$ be the minority/majority
function over $d$ variables. Let $\mathcal{I}(Maj_d)=\mathcal{I}(Min_d)$ be the
influences of the functions $Maj_d,Min_d$. Note that $Min_d$ is the
indicator function of the $H_{1/2}$. Let $f:Q_d \rightarrow \{0,1\}$ be the indicator function of the set
$V$. Then
\begin{eqnarray*}
A(k,d) &=& k(|V \cap  H_{1/2}| - |V -  H_{1/2}|) - E[V, V_d \backslash
V] \\
&=& \frac{2^{d'}}{2} \biggr( k\langle f,Min_{d'} \rangle) -
\mathcal{I}(f) \biggr)
\end{eqnarray*}
We now show the following inequality on Boolean functions over the cube, which
proves the lemma when $k \leq \frac{\pi}{2}\mathcal{I}(Min_d) =
\Oh(\sqrt{d})$.

\begin{claim}
There exists a constant $\alpha < 1$, such that for any Boolean
function $f$ on the $d$-dimensional cube, 
$$ \frac{\pi}{2}\cdot I(Min_d)\cdot \left( \langle f,Min_d \rangle - \alpha \right) \leq
\mathcal{I}(f) $$
  \end{claim}
\begin{proof}
To prove the claim we use some well known facts about Fourier
coefficients of Boolean functions. Note that for a Boolean Function $f:\{0,1\}^d
\rightarrow \{+1,-1\}$ there is a well known and studied change of bases called
the Walsh-Fourier transform. For any subset $S \subseteq [d]$ let $\hat{f(S)}$
be the corresponding Fourier coefficient of $f$. Following are some standard
facts about the Fourier coefficients proofs of which can be found in\cite{ROD}.
\begin{itemize}
  \item $|\hat{Min_d}(\{i\})| \sim \sqrt{\frac{2}{\pi d}}$
  \item $\sum_{|S|\geq 2} \hat{Min_d}(S)^2 \leq (1-\frac{2}{\pi})$
  \item $\mathcal{I}(Min_d) \sim \sqrt{\frac{2}{\pi}d}$
  \item $\sum_i |\hat{f}(\{i\})| \leq \mathcal{I}(f)$
\end{itemize}

Now
\begin{eqnarray*}
\frac{\pi}{2}\mathcal{I}(Min_d)\langle f,Min_d \rangle  &=&
\frac{\pi}{2}\mathcal{I}(Min_d)\biggr(\sum \hat{f}(S)\hat{Min_d}(S)\biggr) \\
&\leq& \frac{\pi}{2}\mathcal{I}(Min_d)\left(\sum_{|S|=1}
|\hat{f}(S)||\hat{Min_d}(S)| + \sum_{|S|\geq2}
\hat{f}(S)\hat{Min_d}(S)\right) \\
&=& \frac{\pi}{2}\mathcal{I}(Min_d)\left(\sum_{|S|=1}
|\hat{f}(S)|\sqrt{\frac{2}{\pi d}} + \sum_{|S|\geq2}
\hat{f}(S)\hat{Min_d}(S)\right) \\
&\leq& \frac{\pi}{2}\mathcal{I}(Min_d)\left(\sum_{|S|=1}
|\hat{f}(S)|\sqrt{\frac{2}{\pi d}} + \sqrt{\sum_{|S|\geq2}
\hat{f}^2(S)}\sqrt{\sum_{|S|\geq2}\hat{Min_d}^2(S)}\right)
\\
&\leq& \sum_{|S|=1} |\hat{f}(S)| +
\frac{\pi}{2}\mathcal{I}(Min_d)\left(\sqrt{1-\frac{2}{\pi}}\right)
\\
&\leq& \mathcal{I}(f) + 
\frac{\pi}{2}\mathcal{I}(Min_d)\left(\sqrt{1-\frac{2}{\pi}}\right)
\\
\end{eqnarray*}

Putting $\alpha = \sqrt{1-\frac{2}{\pi}}$ proves the claim, and thus
also completes the proof of Lemma~\ref{lemma:combopthammer}.
\end{proof}
\end{proof}

\section{Proof of Lemma \ref{lemma:gap_preservation}}\label{sec:gappreservation}
In this section, we prove the gap preservation lemma
\ref{lemma:gap_preservation}.
To prove the lemma we define the following general operation on \maxlin instances on the cube. 

\begin{definition}[Tensor \maxlin]
Given two \maxlin instances $\Gamma_1:E_{d_1} \rightarrow \{0,1\}$ and
$\Gamma_2:E_{d_2} \rightarrow \{0,1\}$ supported on Hypercubes of dimensions
$d_1$ and $d_2$, define a \maxlin instance $\Gamma_1 \otimes \Gamma_2 :
E_{d_1+d_2} \rightarrow \{0,1\}$ on the hypercube of dimension $d_1 + d_2$   as
follows. For an edge $(v_1,v_2)$ let $i(v_1,v_2)$ be the coordinate on which the
corresponding vectors $\mathbf{v_1},\mathbf{v_2}$ differ. Let
$\mathbf{v_i^{d_1}}$,$\mathbf{v_i^{d_2}}$  be the vector $\mathbf{v_i}$
restricted on the first $d_1$ coordinates and the last $d_2$ coordinates respectively. Then 
\begin{equation*}
   \Gamma_1 \otimes \Gamma_2((v_1,v_2))= \left\{
	\begin{array}{ll}
		\Gamma_1((\mathbf{v_1^{d_1},v_2^{d_1}}))  & \mbox{if }  i(v_1,v_2) \in
		[0,d_1-1]
		\\
		\\
		\Gamma_2((\mathbf{v_1^{d_2},v_2^{d_2}}))  & \mbox{if }  i(v_1,v_2) \in
		[d_1,d_1+d_2-1]		
	\end{array}
\right.
\end{equation*}
\end{definition}

Note that the above tensor product defines an edge according to the first
instance or the second instance depending upon the coordinate along which the
edge crosses. We prove the following lemmas about the above defined tensor
product. 

\begin{lemma}
 \label{lemma:gap_preservation:Comb}
 Let $\Gamma_1$ have combinatorial optimum $\geq \beta_1$ and $\Gamma_2$
 have combinatorial optimum $\geq \beta_2$. Then the combinatorial optimum
 of $\Gamma_1 \otimes \Gamma_2$ is $\geq \frac{d_1\beta_1 + d_2\beta_2}{d_1 +
 d_2}$.
 \end{lemma}

 \begin{lemma}
 \label{lemma:gap_preservation:SDP}
 Let $\Gamma_1$ have SDP optimum $\leq \alpha_1$ and $\Gamma_2$ have SDP
 optimum $\leq \alpha_2$. Then the SDP optimum of $\Gamma_1 \otimes \Gamma_2$ is
 $\leq \frac{d_1\alpha_1 + d_2\alpha_2}{d_1 + d_2}$.
 \end{lemma} 
 
 Note that given any $(\alpha,\beta)$ \maxlin Instance $\Gamma$ on the cube of
 dimension $d$ define $\Gamma_i = \otimes_{1}^i \Gamma $. By lemmas
 \ref{lemma:gap_preservation:SDP} and \ref{lemma:gap_preservation:Comb} we get
 that $\Gamma_i$ is an $(\alpha,\beta)$ Instance on the hypercube of dimension
 $i.d$. This proves lemma \ref{lemma:gap_preservation}.

\begin{proof}[Proof of lemma \ref{lemma:gap_preservation:Comb}]
 We prove the lemma by proving that the instance $\Gamma_1\otimes \Gamma_2$ can
 be partitioned into edge disjoint copies of the instances $\Gamma_1$ and
 $\Gamma_2$. Consider any $d_1+d_2$ dimensional vector. Fix the last $d_2$
 coordinates and vary the first $d_1$ coordinates within the space
 $\{0,1\}^{d_1}$. Note that the vectors generated by the above process naturally
 define a subset of edges of the cube $Q_{d_1+d_2}$. Also note that the subset
 of edges generated is an exact copy of $\Gamma_1$. Therefore repeating the
 above process for all choices of the last $d_2$ coordinates gives us $2^{d_2}$
 edge disjoint copies of $\Gamma_1$ within $\Gamma_1 \otimes \Gamma_2$. Fixing
 the first $d_1$ coordinates and repeating the same line of argument as above we
 get $2^{d_1}$ edge disjoint copies of $\Gamma_2$ within $\Gamma_1 \otimes
 \Gamma_2$. Note that the copies described above form an edge partition of
 $\Gamma_1 \otimes \Gamma_2$. Therefore the combinatorial optimum of the
 instance is 
 \begin{eqnarray*}
 &&\geq \frac{1}{E_{d_1+d_2}} \biggr(2^{d_2}\frac{d_12^{d_1}}{2}\beta_1 +
 2^{d_1}\frac{d_22^{d_2}}{2}\beta_2\biggr) \\
 &&\geq \frac{d_1\beta_1 + d_2\beta_2}{d_1 +
 d_2}
 \end{eqnarray*}
 \end{proof}

 \begin{proof}[Proof of lemma \ref{lemma:gap_preservation:SDP}]
 It is enough to give one SDP solution which has the required value. 
 Let the optimal $SDP$ solution for $\Gamma_1$ be $S_1: V_{d_1} \rightarrow
 \R^{2^{d_1}}$ and for $\Gamma_2$ be $S_2: V_{d_2} \rightarrow
 \R^{2^{d_2}}$. 
 
 Define the following solution $S_1 \otimes S_2: V_{d_1+d_2} \rightarrow
 \R^{2^{d_1+d_2}}$. 
 
 $$S_1 \otimes S_2(\mathbf{v}) = S_1(\mathbf{v^{d_1}}) \otimes
 S_2(\mathbf{v^{d_2}})$$

 It is immediate by the properties of tensor products of vectors that $S_1
 \otimes S_2$ is a valid SDP solution. We now compute the SDP value of $S_1
 \otimes S_2$. Let $E_1 \in E_{d_1 + d_2}$ be the set of edges which go through
 the first $d_1$ coordinates and $E_2 \in E_{d_1 + d_2}$ be the set of edges which go through
 the last $d_2$ coordinates. Note that $|E_1| = \frac{d_1}{2}2^{d_1+d_2}$ and
 $|E_2| = \frac{d_2}{2}2^{d_1+d_2}$. Therefore the SDP value achieved by the
 solution is
 
 \begin{eqnarray*}
 &&\frac{1}{E_{d_1+d_2}}\left(\sum\limits_{(v_1,v_2) \in E_1} \|S_1\otimes
 S_2(v_1) \pm S_1\otimes S_2(v_2)\|^2 + \sum\limits_{(v_1,v_2) \in E_2} \|S_1\otimes S_2(v_1) \pm
 S_1\otimes S_2(v_2)\|^2 \right)\\
 &&=\frac{1}{E_{d_1+d_2}}\left(\sum\limits_{(v_1,v_2) \in E_1}
 \|S_2(\mathbf{v_1^{d_2}})\|^2 \| \|S_1(\mathbf{v_1^{d_1}}) \pm S_2(\mathbf{v_1^{d_1}})\|^2 +
 \sum\limits_{(v_1,v_2) \in E_2} \|S_1(\mathbf{v_1^{d_1}})\|^2 \|
 \|S_2(\mathbf{v_1^{d_2}}) \pm S_2(\mathbf{v_1^{d_2}})\|^2 \right)\\
 &&= \frac{1}{E_{d_1+d_2}}\left(\alpha_1d_12^{d_1+d_2} + \alpha_2d_22^{d_1+d_2}
 \right)\\
 &&= \frac{d_1\alpha_1 + d_2\alpha_2}{d_1 + d_2}
 \end{eqnarray*}
 \end{proof}

\section{Towards solving Unique Games on the Hypercube}\label{sec:cycles}
\def\x{\mathbf{x}}
\def\y{\mathbf{y}}

In this section we propose a candidate algorithm for solving Unique Games on the
Hypercube. Our candidate algorithm is simply augmenting the Goemans Williamson
SDP with appropriate triangle inequalities. We conjecture that the augmented SDP
is strong enough to solve unique games on the Hypercube. In particular we show
that our proposed instance $\Delta(k,d)$ defined in the previous section is
indeed solved by this SDP. The motivation behind our conjecture comes from the
fact which we show next that on a cycle of any length the augmented SDP has at
best a constant gap. This implies in particular that an inconsistent cycle in a
graph acts as a certificate for an unsatisfied edge in the SDP solution as well.
Therefore the property of necessarily having many inconsistent cycles makes
\maxlin instances on a graph solvable by SDP. We end the section by showing
that our instance indeed has a lot of inconsistent cycles and by conjecturing
that in fact any instance on the Boolean Cube satisfies this property. 

We begin by defining our augmented SDP. 

\begin{definition}[\gwplus]
Given a graph $G=(V,E)$ and a \maxlin instance $I:E \rightarrow \{0,1\}$ on
it, let the set of equality edges be $E^{+}$ and the set of inequality edges be
$E^{-}$. The augmented Goemans-Williamson SDP for the instance is defined as 
\begin{alignat*}{2}
    \text{minimize }   & \frac{1}{4|E|}\left( \sum\limits_{(u,v) \in E^{+}}
    \|x_u - x_v\|^2 +  \sum\limits_{(u,v) \in E^{-}} \|x_u +
    x_v\|^2 \right)
    \\
    \\
    &\text{subject to } \;\; \|x_u\|^2 = 1 \;\; (\forall \; u \in V) \\
    &\;\; \|a_i - a_j\|^2 \leq \|a_i - a_k\|^2 + \|a_k - a_j\|^2 \;\; (\forall
    \; i,j,k \in V,a_i = \pm x_i,a_j \pm x_j,a_k \pm x_k ) \\
  \end{alignat*}
\end{definition}

One way in which the \gwplus algorithm improves on the
Goemans-Williamson algorithm is that is takes inconsistent cycles into
account, as is formalised below.

\begin{definition} [inconsistent cycles]
  Let $I$ be a \maxlin instance defined on a graph $G$. A cycle in $G$
  is said to be inconsistent if  no assignment can satisfy all
  edges of the cycle (note that there is always an assignment that satisfies all
  edges of a cycle but one). 
\end{definition}

\begin{theorem}
\label{thm:cycletheorem}
Consider a \maxlin instance $I$ defined on a graph $G=(V,E)$, and suppose that
there are $\epsilon\cdot |E|$ edge-disjoint inconsistent cycles in the
instance. Then given $I$, the value returned by \gwplus is at least
$\epsilon$. 
\end{theorem}

Theorem~\ref{thm:cycletheorem} is well known, but we give a proof for
completeness.
\begin{proof}
  First consider the case where the instance just contains one cycle
  $C$.  If it is consistent, it is easy to see that the SDP achieves a
  value of 0. We now focus on inconsistent cycles.  Let $u_0,u_1
  \ldots u_{n-1}$ be the vertices of the cycle in order and let $u_{n}
  = u_0$. Let $E_i$ be edge connecting $u_i \rightarrow u_{i+1}$ and
  let $C(E_i)$ be defined to be $1$ if there is an equality constraint
  on $E_i$ and $-1$ otherwise.  Define $$sign(i) = \Pi_{j=0}^i
  C(E_i)$$Note that w.l.o.g. $sign(0) = 1$ and $sign(n) = -1$ because the cycle
  is inconsistent. The objective function of \gwplus now is thus

\begin{eqnarray*}
\frac{1}{4}\biggr(\sum\limits_{i=0}^{n} \|sign(i)X_{u_i} -
sign(i+1)X_{u_{i+1}}\|^2 \biggr) &\geq& \frac{1}{4}\biggr(\|sign(0)X_{u_0} -
sign(n)X_{u_n}\|^2\biggr) \\
&=& \frac{1}{4} \|2(X_{u_0})\|^2 \\
&=& 1
\end{eqnarray*} 

The first inequality follows from the triangle inequalities added to \gwplus.
The above implies that \gwplus has no gap on a cycle. 

\medskip Now for a general instance, note that the above implies that
any inconsistent cycle in the given instance must contribute at least
$1$ to the value of the objective function in \gwplus. In particular
if we can find $\epsilon\cdot |E|$ inconsistent edge disjoint cycles in the given
instance we can be assured that the \gwplus optimum is at least
$\epsilon$, as required.
\end{proof}

An interesting question is whether there are instances of \maxlin on
the hypercube which are $\epsilon$ unsatisfiable, and yet there are
not enough disjoint inconsistent cycles that certify the value to be
at least $\Omega(\epsilon)$ . We conjecture that in fact there
are no such instances, and therefore that the \gwplus algorithm gives
a constant approximation algorithm for \maxlin on the hypercube.

\begin{conjecture}
Given a \maxlin instance on the Hypercube $(V_d,E_d)$ such that in any
labeling at least $\epsilon$ fraction of its edges are unsatisfied, then there are at least
$\Omega(\epsilon|E_d|)$ edge disjoint inconsistent cycles in the instance.  
\end{conjecture}

One  motivation behind our conjecture is the presence of a large number
of cycles containing every edge -- each edge is contained in
$(d-1)$ 4-cycles.  At least it is true that for our instance $\Delta(k,d)$, defined
previously, the statement of the conjecture holds. Recall that when
$k\leq \Oh(\mathcal I(Maj_d))$, the combinatorial optimum is $\Theta(\frac kd)$

\begin{theorem}
  \label{thm:conjholdsforinstance} Let $I=\Delta(k,d)$ be the \maxlin
  instance defined in Section~\ref{sec:mainthm}, where $k\leq \mathcal
  \Oh(I(Maj_d))=\Oh(\sqrt d)$. Then there are at least
  $\Omega(\frac{k}{d}\cdot |E|)$ edge disjoint inconsistent cycles in $I$, where $E$ is the set
  of edges in $I$.
\end{theorem}

\begin{proof}
  We  first investigate the number of inconsistent edge disjoint
cycles in our instance between two subcubes of dimension $d-k$. The
inconsistent edge disjoint cycles in the whole instance will just be
their union over all subcubes. 

\paragraph{Edge-disjoint paths. } To find the required cycles, we
first consider a $d-k$ dimensional subcube $Q_{d-k}(\x)$ inside our
instance, and let $H_{1/2}$ be the set of vertices with Hamming weight
$\leq \frac{d-k}{2}$ inside it (we only consider the Hamming weight
relative to the subcube).  We would like to find many edge disjoint
simple paths in the cube $Q_{d-k}$ such that for every vertex $v \in
H_{1/2}$ there are at least $\ell=\Theta(\sqrt d)$ paths that start
from it and end in a vertex outside of $H_{1/2}$, and such that at
most $\ell$ paths end at any one vetrex. Note that if $P$ is a path of
this type, and if $Q$ is taken to be the same path but on a
neighbouring subcube $Q_{d-k}(\sigma_i(\x))$ ($\sigma_i$ flips the
$i$'th bit of $\x$), then the two paths can be joined to create an
inconsistent cycle.

Note that the above problem is equivalent to the following flow
system. Let every edge within $Q_{d-k}$ have capacity $1$, and add a
source $s$ that connects to every vertex $v \in H_{1/2}$ with an edge
of capacity $\ell$ and a target $t$ that connects to every vertex
outside of $H_{1/2}$ with an edge of capacity $\ell$. If this system
has a flow that saturates the edges going out of $s$ and into $t$,
then we can find the needed paths in our instance: that follows since
if such a flow exists there must also be an equivalent integral
flow. Once an integral flow is achieved, it is easy to see that it can
be broken into edge-independent paths inside the subcube. 

To see whether the flow system is satisfiable or not we simply need to check the
whether every $s-t$ cut is flow sufficient. Consider any cut $V \subset V_{d-k}$. Note
that the demand of the cut is $\ell|(|V - H_{1/2}| - |H_{1/2}-V|)|$ and the capacity
of the cut is $E(V,V_{d-k}-V)$. Note that lemma \ref{lemma:combopthammer} implies
that for $\ell \leq \Oh(\sqrt{d})$ the cut is flow sufficient. 

\paragraph{Stitching paths together.} For every path $P=P(\x)$ that we
found in $Q_{d-k}(\x)$, we can take a corresponding path $P(\mathbf
y)$ in any other subcube. We thus have a system of disjoint paths in
the subcubes of our instance. Let us show how to stitch them together
to get edge disjoint cycles. For this purpose, consider the graph $G$
on the subcube $Q_{d-k}(\x)$ which connects two points when they are
connected by one of our chosen paths. $G$ is a bipartite graph, and
because of the way the paths were selected, it is regular and each
vertex has degree $\ell$. It is well known that such a graph can
always be partitioned into $\ell$ matchings (e.g. using Hall's
theorem): this means that we can choose a color $i=i(P)$ for each
path, $i\in \{1,\ldots,\ell\}$, such that no vertex connects to two
paths with the same color. 

Now each path $P(\y)$ in a subcube $Q_{d-k}(\y)$ can be matched to
the similar path $P(\y')$ in $Q_{d-k}(\y')$, where $\y'=\sigma_i(\y)$
and $i=i(P)$ is the index chosen by $P$ (since $\ell\leq k$, also
$i\leq k$). Joining the endpoints of
those paths creates an inconsistent cycle, and it is easy to verify that
this indeed gives a system of edge-disjoint inconsistent cycles in
$\Delta(k,d)$. 

\paragraph{Counting cycles.} As we constructed $\ell\cdot 2^{d-k}$ disjoint
paths in each subcube, and since each cycle consists of two such
paths, the total number of cycles is $\ell\cdot 2^{d-k}\cdot
2^k/2=\Omega(k\cdot 2^d)$. Since the number of edges in $\Delta(k,d)$
is $d\cdot 2^d$, the number of cycles is $\Omega(\frac kd\cdot |E|)$
as required. 
\end{proof}

\begin{remark}
  The use of Lemma~\ref{lemma:combopthammer} in the proof above can be
  replaced by a simple and direct probabilistic argument for
  constructing the disjoint paths.
\end{remark}


\ENDDOC

